\documentclass[prl,twocolumn,showpacs,preprintnumbers,amsmath]{revtex4}
\usepackage{graphicx}% Include figure files
\usepackage{dcolumn}% Align table columns on decimal point
\usepackage{bm}% bold math

\begin{document}

\preprint{}

\title{Field-induced spin density wave in (TMTSF)$_2$NO$_3$}

\author{David Vignolles,  Alain Audouard$^{\dagger}$, Marc Nardone and Luc
Brossard}


\affiliation {Laboratoire National des Champs Magn\'{e}tiques
Puls\'{e}s (UMR CNRS-UPS-INSA 5147), 143 avenue de Rangueil, 31432
Toulouse cedex 4, France }

\author{Sabrina Bouguessa and Jean-Marc Fabre}
\affiliation {\'{E}cole Nationale Sup\'{e}rieure de Chimie de
Montpellier, 8 rue de l'\'{E}cole Normale 34296
  Montpellier cedex 5, France}

\date{\today}

\begin{abstract} Interlayer magnetoresistance of the Bechgaard salt
(TMTSF)$_2$NO$_3$ is investigated  up to 50 teslas under pressures
of a few kilobars. This compound, the Fermi surface of which is
quasi two-dimensional at low temperature, is a semi metal under
pressure. Nevertheless, a field-induced spin density wave is
evidenced at 8.5 kbar above $\sim$ 20 T. This state is
characterized by a drastically different spectrum of the quantum
oscillations compared to the low pressure spin density wave state.

\end{abstract}


\maketitle

Bechgaard salts (TMTSF)$_2$X, where TMTSF stands for
tetramethyltetraselenafulvalene and X is an inorganic anion, have
been widely studied over the past twenty years for their very
complex (pressure, magnetic field, temperature) phase diagrams
that involve quasi-one dimensional (q-1D) metallic, spin density
wave (SDW), superconducting and field-induced spin density wave
(FISDW) states (for a review, see Refs. \cite{Rev}). Nevertheless,
FISDW phenomenon is still attracting experimental \cite{Ma03} and
theoretical \cite{Se03} studies. In the framework of the quantized
nesting model \cite{Qnm} and its latest refinements \cite{Se03},
orbital effects stabilize SDW states in compounds with q-1D Fermi
surface (FS) at low temperature through the increase of the 1D
character of the electronic movement. Furthermore, it has also
been stated that a FISDW can only occur provided the groundstate
is superconducting \cite{Ya87}. In line with both of these
predictions, a FISDW has so far only been observed in Bechgaard
salts with both a superconducting groundstate and a q-1D FS,
\emph{e. g.} in slowly cooled (TMTSF)$_2$ClO$_4$ at ambient
pressure or (TMTSF)$_2$PF$_6$ under pressures of a few kilobars.
In contrast, a FISDW has also been reported in
(DMET-TSF)$_2$AuCl$_2$ which remains a q-1D metal at least down to
42 mK \cite{Bi99}. It should also be mentioned that a
field-induced insulating phase, essentially driven by Pauli
effects, has been evidenced in $\tau$-phase organic metals with
q-2D FS \cite{Br03}. The aim of this paper is to argue that, at
variance with the above theoretical statements, a FISDW state can
be observed in a q-2D semi metal,
namely, (TMTSF)$_2$NO$_3$ under pressure.\\
\indent Among the members of the Bechgaard salts family,
(TMTSF)$_2$NO$_3$ exhibits numerous peculiar features. As the
temperature decreases, an anion ordering (AO) transition with wave
vector (1/2, 0, 0) is observed at ambient pressure at a
temperature T$_{AO}$ $\simeq$ 45 K \cite{Po81, Ba99} which remains
independent of the magnetic field up to at least 36 T \cite{Au95}.
This leads to a q-2D FS with compensated electron and hole tubes
\cite{Gr83}. At lower temperature, a  SDW state with
incommensurate wave vector and imperfect nesting is stabilized
(T$_{SDW}$ = 9.1 K at ambient pressure) \cite{Le93}. Under
pressure higher than $\sim$ 8 kbar, the AO transition is shifted
towards slightly higher temperatures \cite{Ka90} and a metallic
state is stabilized at low temperature. Asymmetrical warping of
the FS along the less conducting direction \emph{c}* has been
inferred from the absence of Yamaji features in the angular
dependence of the magnetoresistance both in the SDW \cite{Ha99}
and metallic \cite{Ka95} states. Unlike other Bechgaard salts, no
superconducting transition has been observed at temperatures down
to 0.5 K \cite{Ka90, Ka95}. As it was expected for a q-2D metal,
no sign of a FISDW has been evidenced in (TMTSF)$_2$NO$_3$ for
magnetic fields up to 30 T \cite{Ka90, Ka95}. The absence of a
FISDW in this salt has also been interpreted on the basis of a
nested 2D excitonic phase \cite{Je93}.\\
\indent As it is the case for other Bechgaard salts,
(TMTSF)$_2$NO$_3$ exhibits fast oscillations of the
magnetoresistance in the SDW state at ambient pressure with
frequency F$_H$ = (248 $\pm$ 5) T. Unlike the other Bechgaard
salts, a second oscillation series with frequency F$_L$ = (63
$\pm$ 2) T is observed. This frequency is very close to the
frequency of the magnetoresistance anomalies linked to the FISDW
cascade in other salts. E. g. frequencies of 76 T and 60 T have
been reported for (TMTSF)$_2$PF$_6$ at 6.9 kbar \cite{Kw81} and  8
kbar \cite{Co89}, respectively. However, the oscillations observed
in (TMTSF)$_2$NO$_3$ have a clear sinusoidal shape \cite{Au94,
Bi94} which is not the case of the magnetoresistance anomalies
linked to the FISDW cascade. F$_H$ and F$_L$ have been attributed
to Shubnikov-de Haas orbits linked to AO- and SDW-induced
compensated electron and hole tubes \cite{Ki96}, respectively,
although this interpretation is still under debate \cite{Vi00}. In
the metallic state at 8 kbar, only one oscillation series with
frequency F = 190 T has been observed at 0.5 K in magnetic fields
lower than 19 T \cite{Ka95}. This latter frequency, which behaves
as expected for a q-2D orbit, has been regarded as arising from
F$_H$ although F$_H$, as well as F$_L$ is thought to increase
under pressure due to the increase of both the first Brillouin
zone area and of the warping of the FS sheets.\\
\indent In the following, it is demonstrated that (i) a
field-induced phase transition is observed above $\sim$ 20 T in
(TMTSF)$_2$NO$_3$, starting from the q-2D metallic state obtained
under a pressure of 8.5 kbar, and (ii) the field-induced phase is
characterized by a spectrum of the oscillatory magnetoresistance
strongly different from that observed in the low pressure SDW
state. Since the high pressure oscillatory spectrum and background
magnetoresistance are very similar to that of (TMTSF)$_2$PF$_6$
under high pressure, it is inferred that this phase transition is a FISDW.\\
\indent Magnetoresistance experiments were performed in pulsed
magnetic field up to 50 T (pulse decay duration 0.18 sec.).
Electrical contacts were made to the crystal using annealed gold
wires of 10 $\mu$m in diameter glued with graphite paste.
Alternating current (2 $\mu$A, 20 kHz) was injected parallel to
the \emph{c}* direction (interlayer configuration). A lock-in
amplifier with a time constant of 100 $\mu$s was used to detect
the signal across the potential leads. As reported in Ref.
\cite{Bi93}, and contrary to other Bechgaard salts, the largest
faces of most of the crystals are not perpendicular to the
\emph{c}* direction. For this reason, the ambient pressure
transverse magnetoresistance was first measured with the current
injected along the most conductive direction (\emph{a} axis) at a
temperature of 4.2 K, using a rotating sample holder in order to
determine the direction of \emph{c}*. Subsequent measurements of
the interlayer magnetoresistance were performed in the temperature
range from 1.6 K to 11 K, under pressures up to 8.5 kbar in an
anvil cell \cite{Na01}. Prior to these latter experiments, the
pressure dependence of the interlayer resistance was determined in
a pressure clamp at room temperature [d(ln(R))/dP = -0.11
kbar$^{-1}$] with a manganin piezo resistive sensor. This
parameter was used to determine the pressures achieved in the
anvil cell. As reported in Ref. \cite{Na01}, it is expected that
pressure variation during cooling is very small, owing to the cell
geometry.

\begin{figure}
\centering \resizebox{\columnwidth}{!}{\includegraphics*{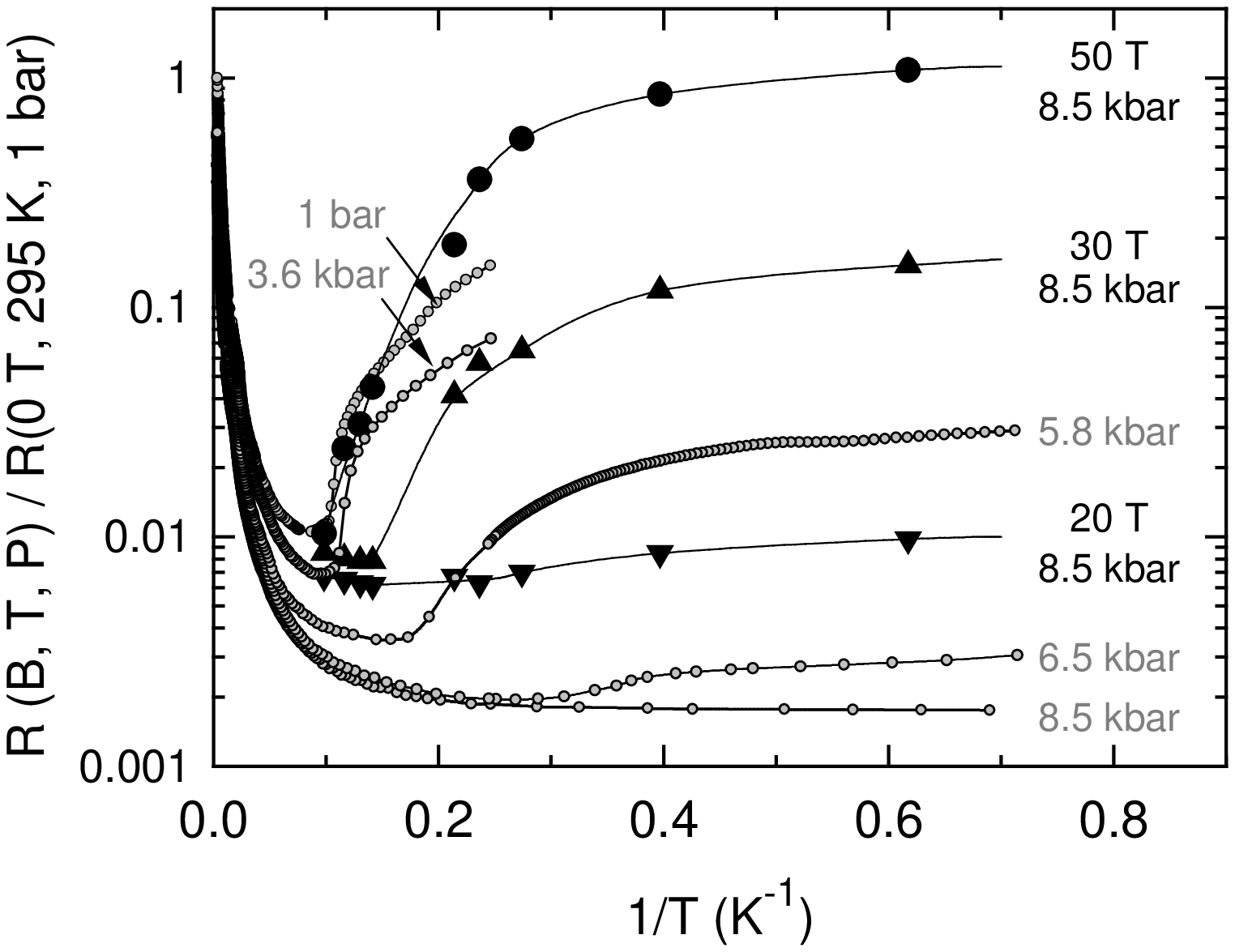}}
\caption{\label{RT} Temperature dependence of the normalized
resistance of (TMTSF)$_2$NO$_3$ at different pressures in
zero-field (small symbols) and in finite magnetic field at 8.5
kbar (large symbols). Solid lines are guides to the eye. }
\end{figure}

\begin{figure}                              %Figure 2
\centering \resizebox{\columnwidth}{!}{\includegraphics*{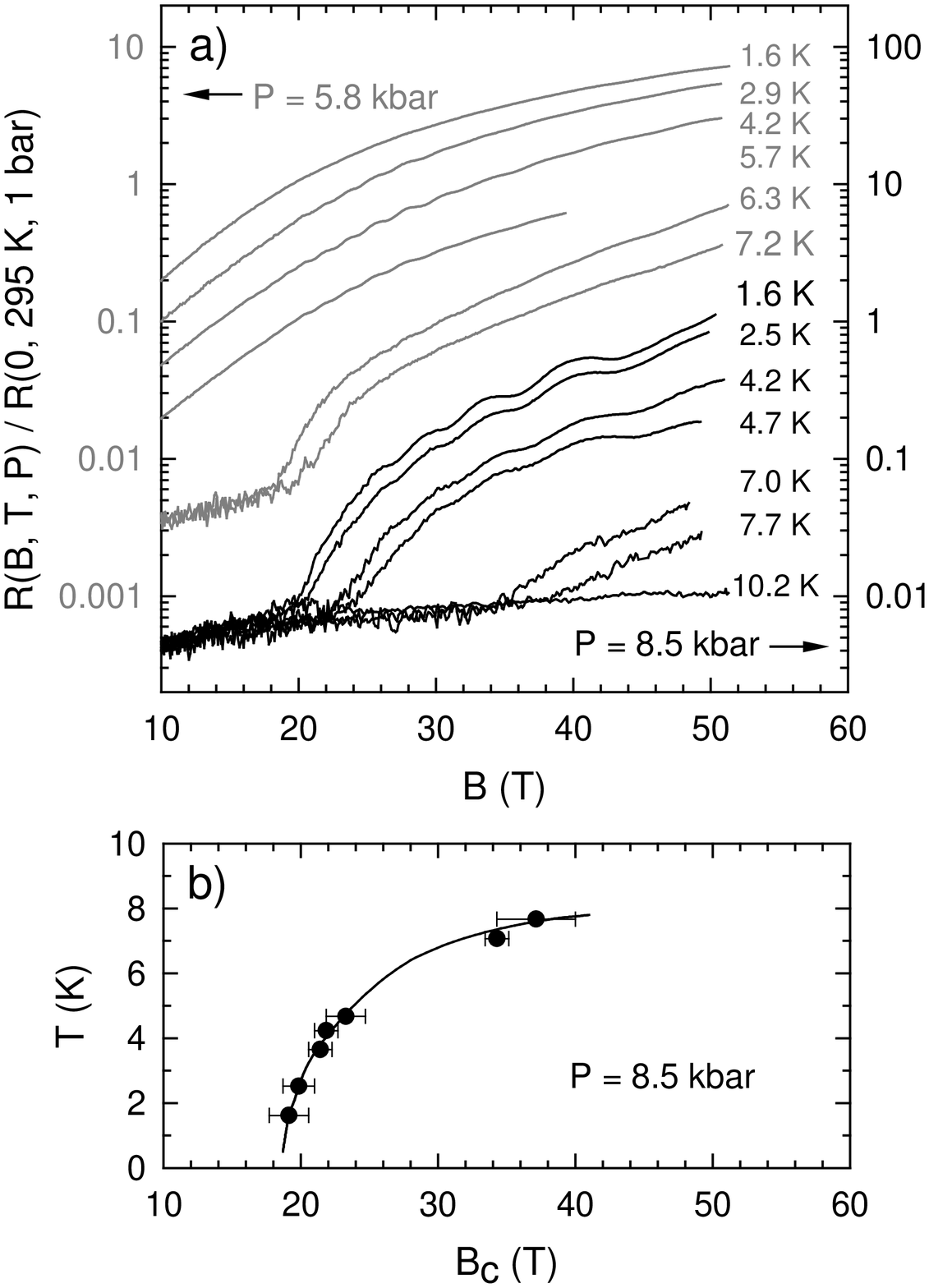}}
\caption{\label{mm} (a) Magnetic field-dependent resistance of
(TMTSF)$_2$NO$_3$ at different temperatures for P = 5.8 kbar (grey
lines) and P = 8.5 kbar (black lines) and (b) semi metal-FISDW
phase transition deduced from the data at P = 8.5 kbar. The solid
line is a guide to the eye.}
\end{figure}

\indent The temperature dependence of the normalized interlayer
resistance is displayed in Fig. \ref{RT}. As previously observed
for in-plane measurements \cite{To89}, the zero field resistance
at ambient pressure exhibits a steep rise below 9.5 K related to
the onset of the SDW transition. In the lower temperature range,
the resistance tends to saturate due to the imperfect nesting. As
the pressure increases, the resistance rise is less and less
pronounced and shifts towards low temperatures. Finally, at 8.5
kbar, the resistance displays a metallic behavior down to the
lowest temperature explored. The AO transition is preserved under
pressure, in agreement with findings of Ref. \onlinecite{Ka90}.
Indeed, T$_{AO}$ deduced from these data increases from 43 K at
ambient pressure up to 50 K at 8.5 kbar. The absence of
superconductivity has been confirmed in a different run performed
in a dilution refrigerator down to 64 mK.

\begin{figure}                                  % Figure 3 (TF)
\centering \resizebox{\columnwidth}{!}{\includegraphics*{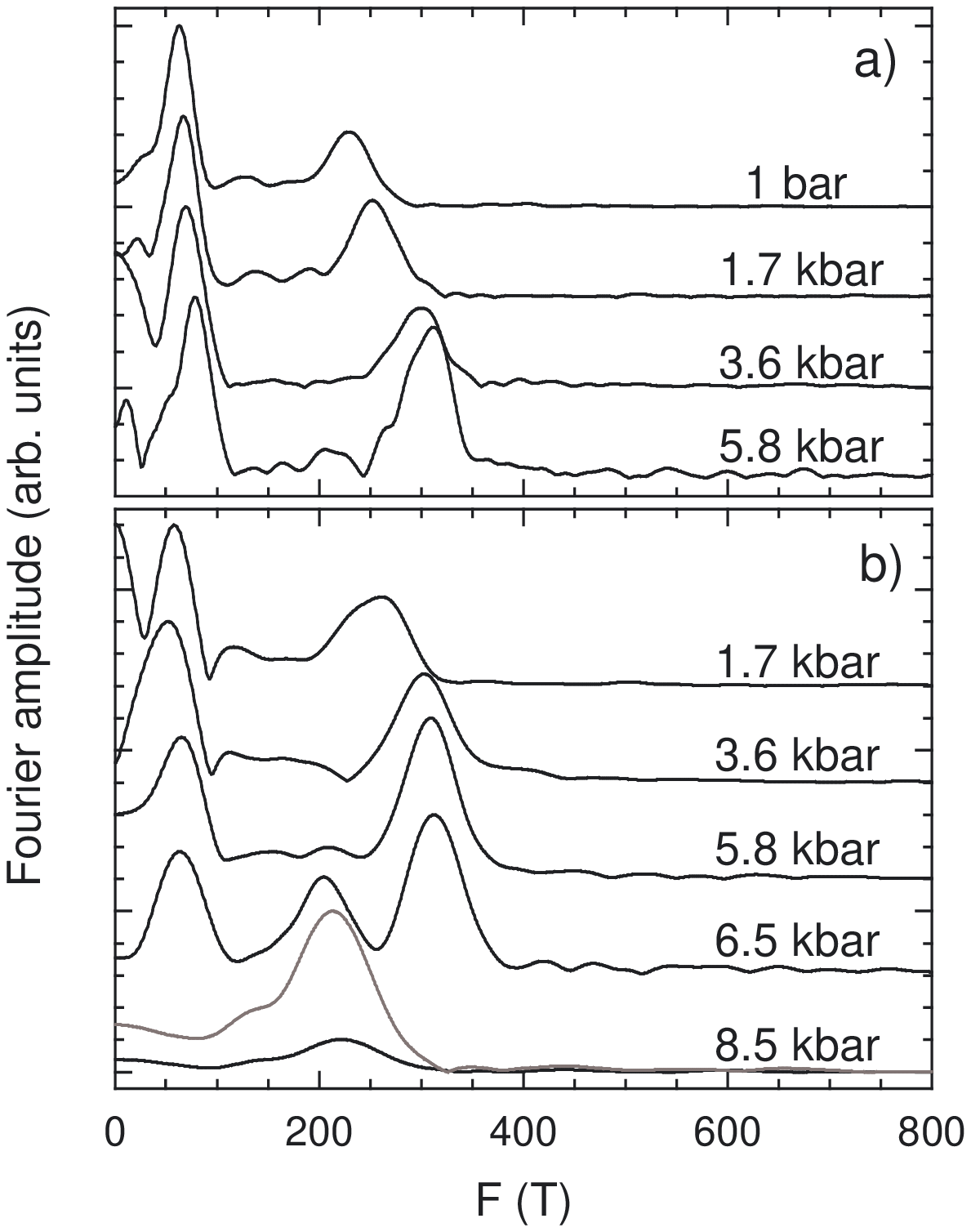}}
\caption{\label{TF} Fourier analysis of the oscillatory
magnetoresistance at different pressures. The black and grey lines
correspond to data at 4.2 K and 1.6 K, respectively. The magnetic
field range is 10 - 36 T in Fig. \ref{TF}(a). In Fig. \ref{TF}(b),
the magnetic field range is 15 T - 50 T for pressures up to 6.5
kbar. For P = 8.5 kbar, the magnetic field range is 22 T - 50 T
(since no oscillation can be detected below 22 T). }
\end{figure}

\begin{figure}                                 % Figure 4 (frequencies = f(P))
\centering
\resizebox{\columnwidth}{!}{\includegraphics*{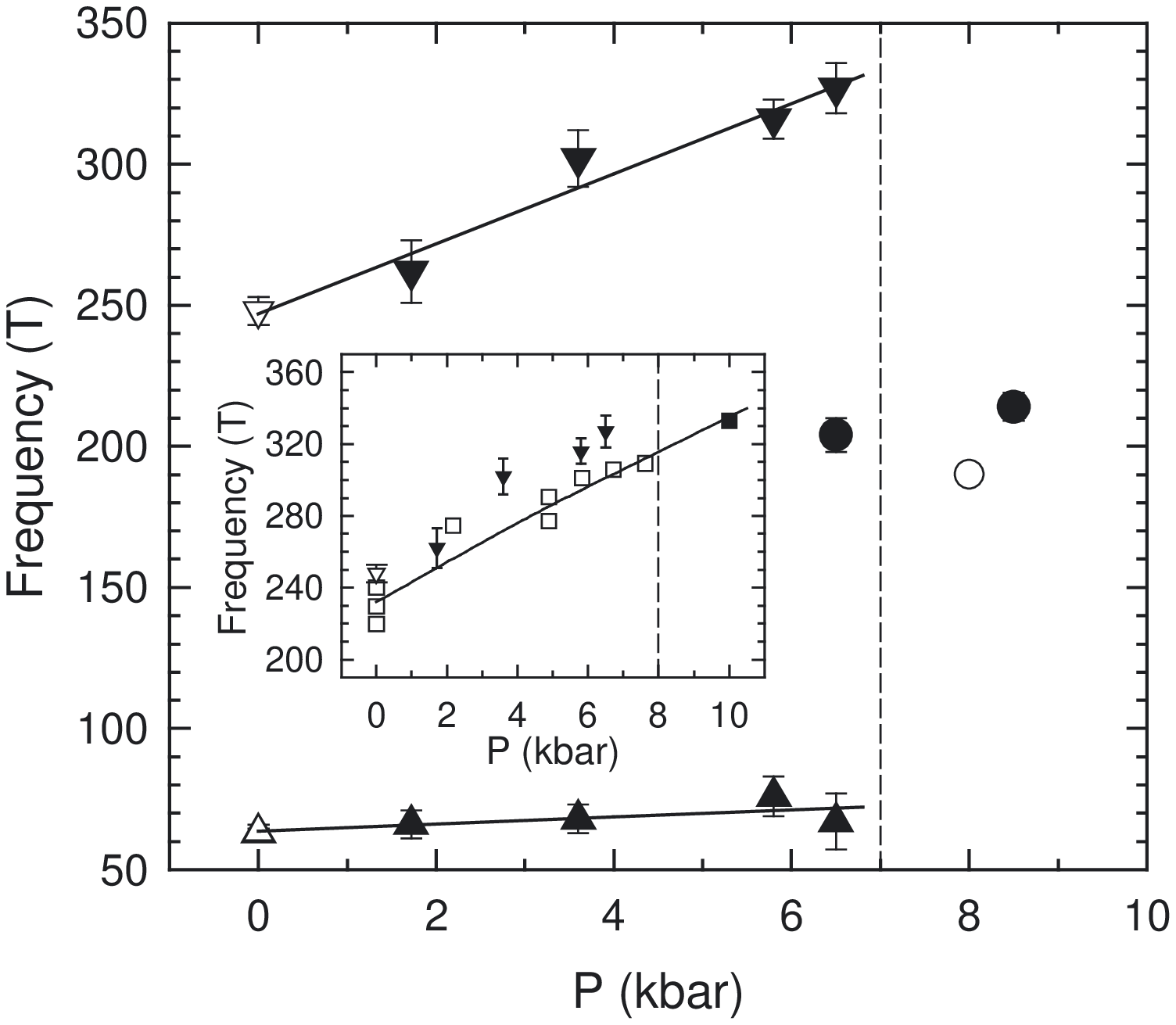}}
\caption{\label{F(P)} Pressure dependence of the frequencies F$_H$
(down triangles), F$_L$ (up triangles) and F$_c$ (circles)
discussed in the text. Solid symbols are deduced from Fig.
\ref{TF}. Open triangles and circle are from Refs. \cite{Au94}
(ambient pressure SDW state) and \cite{Ka95} (metallic state),
respectively. The inset compares the pressure dependence of the
frequency F$_H$ to the pressure dependence of the fast
oscillations in (TMTSF)$_2$PF$_6$. Open squares (SDW state) and
solid square (FISDW state) are from Refs. \cite{Ma01} and
\cite{Ma03}, respectively. The dotted lines mark the transition
between SDW and metallic (in the main frame) or superconducting
(in the inset) groundstates. Solid lines are guides to the eye.}
\end{figure}

\indent Magnetoresistance data collected at different temperatures
are displayed in Fig. \ref{mm}(a). For P = 8.5 kbar, a sudden
resistance rise is observed at temperatures below 10 K at a
threshold field B$_c$ that increases as the temperature increases
[see Fig. \ref{mm}(b)]. This behavior is typical of the resistance
rise due to FISDW transition to the N = 0 state as it is observed
in (TMTSF)$_2$PF$_6$ \cite{Ma03} and (TMTSF)$_2$ClO$_4$
\cite{Ok00, Uj01}. In that respect, it should be noticed that no
FISDW cascade can be detected in the temperature range explored.
The temperature dependence of the resistance at 8.5 kbar and in
magnetic fields above 20 T (see solid symbols in Fig. \ref{RT}) is
also consistent with a FISDW at low temperature. Similar behavior
of the background magnetoresistance is observed at 5.8 kbar, but
at 6.3 K and 7.2 K only, i. e. at temperatures sufficiently above
the zero field SDW transition (T$_c$ = 5.3 K). Nevertheless, as
reported hereafter, the spectrum of the oscillatory
magnetoresistance is
strongly different from the high pressure state.\\
\indent Fourier analysis of the magnetoresistance data at various
pressures is displayed in Fig. \ref{TF}. As expected, F$_H$
increases as the pressure increases. Namely, d[ln(F$_H$)]/dP
$\simeq$ 0.05 kbar$^{-1}$, which is very close to the data for the
fast oscillations in the SDW state of (TMTSF)$_2$PF$_6$
\cite{Ma01} (see inset of Fig. \ref{F(P)}). F$_L$ also increases
with pressure although with a lower rate (d[ln(F$_L$)]/dP $\simeq$
0.02 kbar$^{-1}$, see Fig. \ref{F(P)}). At P = 8.5 kbar, for which
a metallic groundstate is achieved, a drastically different
behavior is observed since only one frequency F$_c$ = (214 $\pm$
5) T is observed in the Fourier spectrum above B$_c$. This feature
is a strong indication that a phase different from the low
pressure SDW phase is induced above B$_c$.

\indent It should be noted that the observed pressure dependence
of the oscillatory spectrum is at variance with the behavior of
(TMTSF)$_2$PF$_6$ and (TMTSF)$_2$ClO$_4$ for which only one
frequency is observed whatever the pressure is. In the former
salt, the oscillation frequency increases monotonously as the
pressure increases so that no significant disruption between SDW
and FISDW states is observed (see inset of Fig. \ref{F(P)}). In
the latter salt, depending on the considered FISDW subphase,
either one or two out of phase series with the same frequency are
observed in the FISDW state \cite{Ok00, Uj01} while only one
series is observed in the metallic state \cite{Uj97}. In short,
whereas the oscillatory spectrum of (TMTSF)$_2$PF$_6$ and
(TMTSF)$_2$ClO$_4$ does not evidence abrupt changes in the
different (metallic, SDW or FISDW) phases, a drastic
pressure-induced modification of the spectrum is observed in
(TMTSF)$_2$NO$_3$. Besides, the oscillation amplitude (A) at 8.5
kbar continuously increases as the temperature decreases and
follow the LK behavior. This is at variance with the behavior
observed in the FISDW state of (TMTSF)$_2$ClO$_4$ \cite{Ok00,
Uj97, Uj01, Qu00} and in the SDW state of (TMTSF)$_2$NO$_3$ for
which A goes to a maximum at a temperature of a few K. The present
study does not display any oscillation in the metallic state down
to 1.6 K and below B$_c$. Nevertheless, it can be remarked that
F$_c$ is close to the frequency (F = 190 T) that has been reported
in the metallic state below 20 teslas, at a pressure of 8 kbar but
at a lower temperature (T = 0.5 K) \cite{Ka95}. Finally, it is
interesting to note that, at the pressure of 6.5 kbar, F$_L$,
F$_H$ and F$_c$ are observed simultaneously in the oscillatory
spectrum (see Fig. \ref{TF}). This may be the signature of some
precursor effects,
rather than phase mixing induced by some pressure inhomogeneity.\\
\indent In conclusion, at variance with the idea that a FISDW
state can only be observed in q-1D superconductors, we have
evidenced a FISDW state in a q-2D semi metal. As demonstrated by
the measured pressure dependence of the oscillatory spectrum, the
(presumably N = 0) FISDW state is strongly different from the
ambient pressure SDW state. A new model, explaining FISDW in q-2D
metals is clearly
needed. \\

\begin{acknowledgments}
We are indebted to H\'{e}l\`{e}ne Bouchiat, Sophie Gu\'{e}ron and
Richard Deblock for having made available their dilution
refrigerator and for their assistance.  We wish to acknowledge
Geert Rikken and Denis J\'{e}rome for stimulating discussions.
\end{acknowledgments}

\hspace{1pc}

 $\dagger$ E-mail: audouard@lncmp.org

 \bibliography{apssamp}% Produces the bibliography via BibTeX.

\begin{references}

\bibitem[1]{Rev} \emph{e. g. Low Dimensional Conductors and Superconductors}, edited by D. J\'{e}rome and L. G.
Caron, NATO Advanced Studies Institute, Series B, Vol. 155
(Plenum, New York, 1987) and T. Ishiguro, K. Yamaji and G. Saito,
\emph{Organic Superconductors}, 2nd ed. (Springer Verlag,
Heidelberg, 1998).

\bibitem[2] {Ma03} N. Matsunaga, K. Yamashita, T. Oota, K.
Nomura, T. Sasaki, T. Hanajiri, J. Yamada and S. Nakatsuji,
Physica B \textbf{329-333} 1154 (2003).

\bibitem[3]{Se03} K. Sengupta and N. Dupuis, Phys.
Rev. B \textbf{68} 094431 (2003).

\bibitem[4]{Qnm} L. P. Gorkov and A. G. Lebed, J. Physique Lett.
\textbf{45} L-433 (1984); M. H\'{e}ritier, G. Montambaux and P.
Lederer, J. Physique Lett. \textbf{45} L-943 (1984).

\bibitem[5]{Ya87} V. M. Yakovenko, Zh. Eksp. Teor. Fiz. \textbf{93} 627 (1987) [Soviet Physics JETP \textbf{66} 355
(1987)]; V. M. Yakovenko, Fizika \textbf{21} Supp.3 44 (1989).

\bibitem[6]{Bi99}N. Bi\v{s}kup, J. S. Brooks, R. Kato and K. Oshima, Phys. Rev. B \textbf{60}
R15005 (1999).

\bibitem[7]{Br03} J. S. Brooks, D. Graf, E. S. Choi, L. Balicas,
K. Storr, C. H. Mielke and G. C. Papavassiliou, Phys. Rev. B
\textbf{67} 153104 (2003).

\bibitem[8]{Po81} J. P. Pouget, R. Moret, R. Comes and K. Bechgaard, J. Physique Lett. \textbf{42} 543 (1981).

\bibitem[9]{Ba99}Y. Barrans, J. Gaultier, S. Brachetti,  P. Guionneau, D. Chasseau and J. M. Fabre, Synth. Met.
\textbf{103} 2042 (1999).

\bibitem[10]{Au95} A. Audouard and S. Ask\'{e}nazy, Phys. Rev. B \textbf{52} 700 (1995).

\bibitem[11]{Gr83}P. M. Grant, Phys. Rev. Lett. \textbf{50} 1005
(1983); S. H\'{e}brard-Brachetti, PhD thesis, Bordeaux (1996).

\bibitem[12] {Le93} L. P. Le, A. Keren, G. M. Luke, B. J.
Sternlieb, W. D. Wu, Y. J. Uemura, J. H. Brewer, T. M. Riseman, R.
V. Upasani, L. Y. Chiang, W. Kang and P. M. Chaikin, Phys. Rev. B.
\textbf{48} 7284 (1993).

\bibitem[13]{Ka90} W. Kang, S. T. Hannahs, L. Y. Chiang and P. M. Chaikin, Phys. Rev. Lett. \textbf{65} 2812
(1990).

\bibitem[14]{Ha99} H. I. Ha, W. Kang and J. M. Fabre, Synth. Met. \textbf{103} 2117 (1999)


\bibitem[15]{Ka95} W. Kang, K. Behnia, D. J\'{e}rome, L.
Balicas, E. Canadell, M. Ribault and J. M. Fabre, Europhys. Lett.
\textbf{29} 635 (1995).

\bibitem[16] {Je93} D. J\'{e}rome, C. R. Acad. Sci. Paris, \textbf{317} S\'{e}r. II 569
(1993).

\bibitem[17]{Kw81} J. F. Kwak, J. E. Shirber, R. L. Greene and E. M. Engler, Phys. Rev. Lett. \textbf{46} 1296 (1981).

\bibitem[18]{Co89} J. R. Cooper, W. Kang, P. Auban, G. Montambaux, D.  J\'{e}rome and K. Bechgaard, Phys. Rev. Lett.
\textbf{63} 1984 (1989).

\bibitem[19]{Au94} A. Audouard, F. Goze, S. Dubois, J. P. Ulmet, L.
Brossard, S. Ask\'{e}nazy, S. Tomi\'{c} and J. M. Fabre, Europhys.
Lett. \textbf{25} 363 (1994); A. Audouard, F. Goze, J. P. Ulmet,
L. Brossard, S. Ask\'{e}nazy and J. M. Fabre, Phys. Rev. B
\textbf{50} 12726 (1994).

\bibitem[20]{Bi94}N. Bi\v{s}kup, L. Balicas, S. Tomi\'{c}, D. J\'{e}rome and J. M. Fabre, Phys. Rev. B \textbf{50}
12721 (1994).

\bibitem[21]{Ki96}K. Kishigi and K. Machida, Phys. Rev. B \textbf{53} 5461
(1996); K. Kishigi and K. Machida, J. Phys-Condens. Matter
\textbf{9} 2211 (1997).

\bibitem[22]{Vi00} D. Vignolles, J. P. Ulmet, A. Audouard, M. Naughton and J. M. Fabre, Phys. Rev. B
\textbf{61} 8913 (2000).

\bibitem[23]{Bi93}N. Bi\v{s}kup,M. Basleti\'{c}, S. Tomi\'{c}, B.
Korin-Hamzi\'{c},K. Bechgaard and J. M. Fabre, Phys. Rev. B
\textbf{47} 8289 (1993).

\bibitem[24] {Na01} M. Nardone, A. Audouard, D. Vignolles and L.
Brossard, Cryogenics \textbf{41} 175 (2001).

\bibitem[25] {To89} S. Tomi\'{c}, J. R. Cooper, D. J\'{e}rome and K. Bechgaard,  Phys. Rev. Lett. \textbf{62} 462 (1989);
S. Tomic, J. R. Cooper, W. Kang, D. J\'{e}rome and K. Maki, J.
Phys. (France) I \textbf{1} 1603 (1991).

\bibitem[26] {Ok00}Ok-Hee Chung, W. Kang, D. L. Kim and C. H.
Choi, Phys. Rev. B \textbf{61} 11649 (2000).


\bibitem[27] {Uj01} S. Uji, C. Terakura, T. Terashima, J. S.
Brooks, S. Takasaki, S. Maki, S. Nakatsuji and J. Yamada, Current
Applied Physics \textbf{1} 77 (2001).

\bibitem[28] {Ma01} N. Matsunaga, K. Yamashita, H. Kotani, K.
Nomura, T. Sasaki, T. Hanajiri, J. Yamada,  S. Nakatsuji and H.
Anzai, Phys. Stat. Sol (b) \textbf{223} 545 (2001).


\bibitem[29] {Uj97} S. Uji, J. S. Brooks,  S. Takasaki, J. Yamada
and H. Anzai, Solid State Commun. \textbf{103} 387 (1997).

\bibitem[30] {Qu00} J. S. Qualls, C. H. Mielke, J. S. Brooks, L.
K. Montgomery, D. G. Rickel, N. Harrison and S. Y. Han, Phys. Rev.
B \textbf{62} 12680 (2000).

\end{references}

 \end{document}